\documentclass[conference]{IEEEtran}
\IEEEoverridecommandlockouts

\usepackage{hyperref}
\usepackage{url}
\usepackage{cite}
\usepackage{amsmath,amssymb,amsfonts}
\usepackage{algorithmic}
\usepackage{graphicx}
\usepackage{textcomp}
\usepackage{xcolor}
\usepackage{comment}
\usepackage{multirow}
\usepackage{cite}
\usepackage{subcaption}
\usepackage{booktabs}
\usepackage{array}
\usepackage{tikz}
\usepackage{caption}

\usepackage{cleveref}
\usepackage{booktabs}
\usepackage{arydshln}
\usepackage{enumitem}
\usepackage{epstopdf}
\usepackage{makecell}

\crefname{figure}{Fig.}{Figs.}
\Crefname{figure}{Fig.}{Figs.}
\crefname{table}{Tab.}{Tabs.}
\Crefname{table}{Tab.}{Tabs.}
\crefname{equation}{Eq.}{Eqs.}
\Crefname{equation}{Eq.}{Eqs.}
\crefname{section}{Section}{Secs.}
\Crefname{section}{Section}{Secs.}
\crefname{appendix}{Appendix}{Appendices}
\Crefname{appendix}{Appendix}{Appendices}

\def\BibTeX{{\rm B\kern-.05em{\sc i\kern-.025em b}\kern-.08em
    T\kern-.1667em\lower.7ex\hbox{E}\kern-.125emX}}

\begin{document}

\title{SenSE: Semantic-Aware High-Fidelity Universal Speech Enhancement}


\author{
\IEEEauthorblockN{
Xingchen Li$^{1}$, Hanke Xie$^{1}$, Ziqian Wang$^{1}$, 
Zihan Zhang$^{2}$, Longshuai Xiao$^{2}$, Shuai Wang$^{3}$, Lei Xie$^{1*}$\thanks{*Corresponding author.}
}
\IEEEauthorblockA{
$^{1}$Audio, Speech and Language Processing Group (ASLP@NPU), \\
School of Computer Science, Northwestern Polytechnical University, China \\
$^{2}$Huawei Technologies Co., Ltd., China \\
$^{3}$Nanjing University, China
}
}

\maketitle

\begin{abstract}
Generative Universal Speech Enhancement (USE) methods aim to leverage generative models to improve speech quality under various types of distortions. However, existing generative speech enhancement methods often suffer from semantic inconsistency in the generated outputs. Therefore, we propose SenSE, a novel two-stage generative universal speech enhancement framework, by modeling semantic priors with a language model, the flow matching-based speech enhancement process is guided to generate semantically faithful speech, thereby effectively improving context fidelity. In addition, we introduce a dual-path masked conditioning training strategy that enables flow matching-based enhancement to flexibly integrate multi-source conditioning signals from degraded speech, semantic tokens, and reference speech, thereby improving model flexibility and adaptability. Experimental results demonstrate that SenSE achieves state-of-the-art performance among generative speech enhancement models and exhibits a high performance ceiling, particularly under challenging distortion conditions. Codes and demos are available at \url{https://github.com/ASLP-lab/SenSE}.
\end{abstract}


\section{Introduction}
\label{sec:intro}

Speech enhancement (SE) is designed to improve both the perceptual quality and intelligibility of speech degraded under adverse acoustic conditions, such as background noise, reverberation, signal clipping, and bandwidth constraints. Most prior approaches are designed to handle a single type of distortion, whereas Universal Speech Enhancement (USE) methods that employ a single model to address multiple distortion types have recently attracted increasing attention ~\cite{liu2022voicefixer, zhang2025anyenhance, kang2025llase, zhang2025composite}. Discriminative models ~\cite{hu2020dccrn, fu2022uformer, rong2024gtcrn} typically establish a direct mapping from degraded speech to clean speech. However, they tend to introduce additional distortion and artifacts under severely degraded conditions, which can substantially degrade perceptual quality.

Recent generative speech enhancement approaches have demonstrated strong potential for producing more natural and perceptually coherent speech ~\cite{tai2023dose, wang2024selm, yao2025gense, zhang2025anyenhance}. These approaches capture the distribution of clean speech and generate the corresponding clean signal when conditioned on degraded input. While such approaches can produce high-quality speech, their major limitation lies in the difficulty of ensuring content fidelity. In practice, a substantial amount of speech hallucination still occurs, which makes it challenging to preserve the original speech structure.

Unlike other generative tasks, speech enhancement places greater emphasis on leveraging generative models to reconstruct high-fidelity clean speech. However, due to the high complexity of speech distortions, distinguishing speech components from interfering signals can become particularly challenging in certain scenarios. This challenge constitutes a fundamental limitation of many generative speech enhancement approaches: inaccurate identification of speech components in degraded speech often leads to a mismatch between the generated output and the ground-truth clean speech, thereby reducing reconstruction fidelity. This degradation is typically reflected in lower semantic similarity and speaker similarity. Some existing approaches generate discrete tokens using language models \cite{yao2025gense, kang2025llase} or masked generative models (MGMs) \cite{zhang2025anyenhance}. However, the inherent information loss of discrete representations, together with the difficulty of predicting fine-grained acoustic tokens, prevents these methods from effectively resolving semantic inconsistency. Diffusion- or flow-based methods generally establish a more direct mapping between the distributions of degraded and clean speech, and thus often achieve higher overall fidelity. Nevertheless, the issue of speech component confusion persists, especially under severe distortion conditions.

To address the above challenge, we provide semantic prior guidance to flow matching-based speech enhancement models. Recent advances in speech tokenizers ~\cite{du2024cosyvoice} have demonstrated that language-related semantic information can be encoded into discrete semantic tokens, providing a reliable intermediate representation for leveraging semantic priors in generative speech enhancement. Building upon this insight, we aim to explicitly model the distribution of semantic tokens directly from complete degraded speech inputs, for which the strong contextual modeling capability of language models offers a promising solution. Unlike prior language-model-based speech enhancement approaches, we leverage language models to model semantic priors rather than to directly generate complete speech.

\begin{figure*}[ht]
  \centering

  \includegraphics[width=0.66\textwidth]{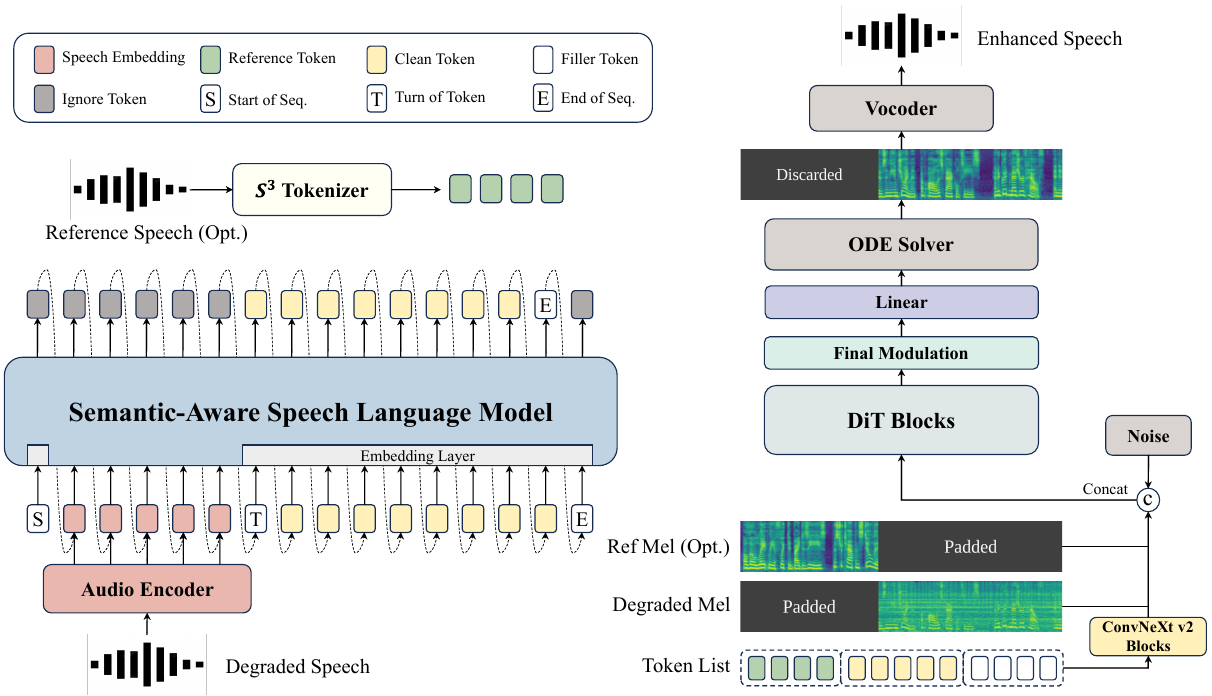}
  \caption{An overview of the two-stage architecture in SenSE with explicit semantic modeling. }
  \label{fig:overall}
  
\end{figure*}

In this paper, we propose SenSE, a novel speech enhancement framework that leverages a language model to explicitly model semantic information from speech and introduces a semantic guidance mechanism that injects semantic tokens into flow matching-based speech enhancement. Moreover, we introduce a dual-path masked conditional training strategy. This design enables the model to fully exploit multiple cues for speech enhancement, including degraded speech, semantic tokens, and reference speech. In particular, the model can selectively leverage high-quality reference speech from the same speaker to compensate for the loss of speaker-related information under severe distortion conditions. Experimental results demonstrate that the proposed approach achieves performance comparable to that of current large-scale generative speech enhancement models, even with a relatively small model size and low computational overhead. Furthermore, by appropriately scaling up the model capacity, the framework exhibits strong robustness and performance gains in severely distorted scenarios.

\section{Method}

\label{headings}


As illustrated in \Cref{fig:overall}, SenSE is organized into two key stages: 1) \textit{semantic-aware speech language model}: this module adopts a language model to derive purified semantic tokens from the continuous speech embeddings obtained by encoding degraded speech. 2) \textit{semantic-guided speech enhancement with flow matching}: the semantic tokens obtained from the first stage are incorporated as additional conditioning in flow matching-based speech enhancement, thereby enhancing the preservation of semantic information during generation. The dual-path masked conditional training strategy enables the model to fully leverage all available information to generate clean speech.

\subsection{Semantic-Aware Speech Language Model}
\label{sec:saslm}

To enhance robustness under conditions of severe distortion, we introduce a Semantic-Aware Speech Language Model (SASLM). We adopt a randomly initialized LLaMA model as our backbone and employ an audio encoder to extract continuous speech representations as input to the model.

We adopt the $ \mathcal{S}^3 $ Tokenizer as the speech tokenizer. Originally introduced in CosyVoice ~\cite{du2024cosyvoice}, it integrates a quantization module, either vector quantization (VQ) ~\cite{van2017neural} or finite scalar quantization (FSQ) ~\cite{mentzer2023finite}, into the encoder of the SenseVoice ~\cite{an2024funaudiollm} ASR model to generate discrete tokens under supervised training.

During training, the input sequence to SASLM is formatted as “\textless Start of Seq. \textgreater, speech embedding, \textless Turn of Token \textgreater, semantic token, \textless End of Seq. \textgreater”, where \textless Start of Seq. \textgreater denotes the start-of-sequence token, \textless Turn of Token \textgreater indicates the transition from speech embeddings to speech tokens, and \textless End of Seq. \textgreater marks the end of the sequence. The model is trained under a next-token prediction objective. The training objective is as follows:
\begin{equation}
    p(x) = \prod_{k=1}^{n} p\bigl(s_k \mid s_1, \ldots, s_{k-1}, e_{1\ldots n})
\end{equation}
where $s$ denotes the semantic tokens of clean speech, $e$ represents the speech embeddings extracted from the degraded speech by the audio encoder, and $n$ indicates the total number of frames in the input speech sequence.

At inference time, a sequence consisting of the start-of-sequence symbol, the speech embeddings, and the turn-of-token indicators is provided to the SASLM as a prefix. The model then generates the semantic tokens corresponding to the clean speech via next-token prediction. In addition, when reference speech is available, it is converted into reference tokens using the $\mathcal{S}^3$ Tokenizer, which are then used in the flow matching stage to facilitate the alignment between semantic and acoustic cues.

\subsection{Semantic-Guided Speech Enhancement with Flow Matching}
\label{sec:fm}

Based on the semantic tokens predicted by SASLM and the reference token (if provided), we incorporate a semantic guidance mechanism into flow matching-based speech enhancement, which constitutes the second stage of our model. With this mechanism, the model jointly leverages acoustic cues from the degraded mel spectrogram and semantic cues from the semantic tokens. In this way, the model generates clean speech conditioned on semantic tokens, while ensuring that the enhanced output preserves a spectral envelope similar to that of the distorted input, thereby maintaining high fidelity. In addition, the proposed dual-path masked conditional training strategy encourages the flow matching model to effectively leverage semantic information rather than relying solely on acoustic cues, while also enabling the optional use of clean reference speech from the same speaker to compensate for speaker-related information loss under severe distortion conditions.

\begin{figure}[t]
    \centering
    \includegraphics[width=0.75\linewidth]{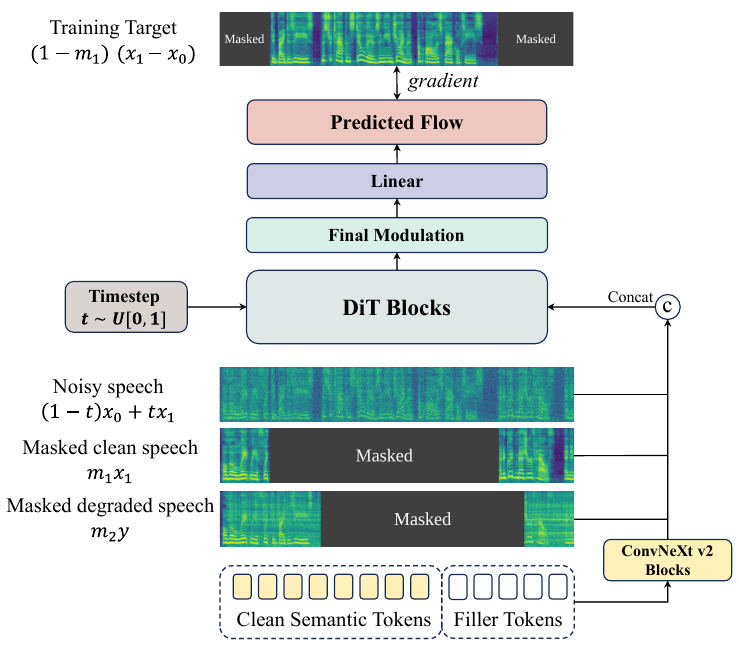}
    \caption{Illustration of the proposed dual-path masked conditional training strategy.}
    \label{fig:cfm_training}
\end{figure}

\textbf{Dual-path masked conditional training}\quad As illustrated in \Cref{fig:cfm_training}, the second-stage flow matching model is trained on a conditioned speech infilling task using a DiT ~\cite{peebles2023scalable} backbone. We adopt DiT blocks following the F5-TTS design ~\cite{chen2024f5}, with a zero-initialized adaptive Layer Normalization (adaLN-zero) module as the final modulation mechanism. During training, triplets of clean speech, degraded speech, and semantic tokens extracted from clean speech are used. The clean and degraded speech mel spectrograms are denoted as $ x_1 \in \mathbb{R}^{F \times T} $ and $ y \in \mathbb{R}^{F \times T} $, respectively, while the semantic token sequence is denoted as $ z $.

\begin{table}[t]
  \centering
  \caption{Constructed distortion types and probabilities in training data simulation.}
  \label{tab:distortion}
  \resizebox{0.95\linewidth}{!}{%
  \begin{tabular}{lcc}
    \toprule
    \textbf{Family} & \textbf{Probability} & \textbf{Hyperparameters} \\
    \midrule
    Noise & 0.9 & SNR $\in [-10, 10]$ \\
    Reverberation & 0.5 & - \\
    Clipping & 0.25 & Clipping threshold $\in [0.05, 0.90]$ \\
    Bandwidth  Limitation & 0.5 & Bandwidth $\in \{2, 4, 8, 16, 22.05\}$kHz \\
    \bottomrule
  \end{tabular}
  }
\end{table}

\begin{table}[t]
  \centering
  \caption{Details of model configurations.}
  \label{tab:model_config}
  \resizebox{0.95\linewidth}{!}{%
  \begin{tabular}{lcccc}
    
    \toprule
    \textbf{Model} & \textbf{Audio Encoder} & \textbf{SASLM} & \textbf{DiT Blocks} & \textbf{Vocoder}  \\
    \midrule
    SenSE$_{base}$ & Whisper-large-v3 & 1536,20,16 & 1024,22,16 & BigVGAN \\
    SenSE$_{small}$ & conformer & 1024,6,16 & 768,18,12 & BigVGAN \\
    SenSE$_{tiny}$ & Whisper-small & 256,16,4 & 768,12,8 & Vocos \\
    \bottomrule
  \end{tabular}
  }
\end{table}

Following the flow matching formulation ~\cite{lipman2022flow}, the model input is constructed as $ (1-t)x_0 + tx_1 $, where $x_0 \sim \mathcal{N}(0, I)$ represents Gaussian noise sampled from the prior distribution, and $ t \sim \mathcal{U}[0,1] $ denotes the sampled flow step. We combine both acoustic cues and semantic cues as the conditioning input to the model. The acoustic cues include the masked clean speech $ m_1 \odot x_{1} $ and the incomplete degraded speech  $ m_2 \odot y $ , where $ m_1 , m_2$ are binary temporal masks. For the semantic cues, semantic tokens are padded with filler tokens to match the length of the mel spectrogram and then encoded by ConvNeXt V2 ~\cite{woo2023convnext} modules to implicitly align the token sequence with the speech features. The condition is constructed by concatenating the clean speech mel-spectrogram, the degraded speech mel spectrogram, and the aligned semantic cues along the feature dimension, based on which the model is trained to reconstruct the masked regions of the clean speech. Therefore, our objective is to train a model $ v_\theta $ to predict the velocity $ v_\theta((1-t)x_0+tx_1, m_1x_1, m_2y, z) $ with target as $ v=(1-m_1)(x_1-x_0) $. We use the mean squared error (MSE) loss, formally represented as:
\begin{equation}
\begin{split}
\mathcal{L}_{\text{FM}}
= \mathbb{E}_{t, x_0, x_1} \Big\|
& v_\theta \big( (1-t)x_0 + t x_1,\; m_1 x_1,\; m_2 y,\; z \big) \\
&\quad - (1-m_1)(x_1 - x_0)
\Big\|^2
\end{split}
\end{equation}
Regarding the rationale for randomly masking the degraded speech, which we refer to as the \textit{degrad mask}, we note that degraded signals often provide direct acoustic cues, which can dominate the learning process. When the model is conditioned simultaneously on degraded speech and semantic tokens, it tends to over-rely on the degraded speech, thereby neglecting the more challenging task of mapping semantic tokens to clean speech. To counteract this bias, we introduce random temporal masking on the degraded speech. This strategy forces the model to reconstruct the masked regions of the degraded speech by leveraging the information provided by the semantic tokens. Through this masking-based training paradigm, the model learns to align semantic tokens with the corresponding mel spectrograms of clean and degraded speech and to more effectively utilize the semantic information.

\textbf{Inference}\quad To generate enhanced speech, we prepare the degraded speech $y_{degraded} \in \mathbb{R}^{F \times T_{1}} $, the semantic tokens $z_{gen}$ predicted by SASLM , and an optional reference speech $x_{ref} \in \mathbb{R}^{F \times T_{2}} $. During inference, as shown in \Cref{fig:overall}, we pad the tail of $x_{ref}$ with empty frames of length $T_1$, which serves as the portion to be generated. Similarly, we pad the front of $y_{degraded}$ with empty frames of length $T_2$ so that it aligns with the generative portion, and the semantic token sequence 
$z_{gen}$, formed by concatenating the reference tokens and the purified tokens, is extended with filler tokens in the same way as in training. When no reference speech is provided, $T_2$ is set to zero, resulting in an all-empty reference channel without requiring additional padding for $y_{degraded}$. The model estimates the velocity field $v$, and the Euler ODE solver is applied to generate the enhanced mel spectrogram. Finally, the mel spectrogram is converted into a waveform using a pretrained vocoder.

\section{Experiment And Results}

\label{others}

\begin{table}[t]
  \centering
  \caption{Model size and RTF of SenSE and other comparison models.}
  \label{tab:model_size}
  \resizebox{0.55\linewidth}{!}{%
  \begin{tabular}{lccc}
    
    \toprule
    \textbf{Model} & \textbf{\#Param} & \textbf{RTF}  \\
    \midrule
    TF-GridNet     & 8.0M    & 0.071  \\
    PGUSE          & 5.1M    & 0.064  \\
    GenSE          & 1092.9M   & 2.031 \\
    LLaSE-G1       & 1895.6M   & 0.049  \\
    FlowSE         & 350.1M    & 0.201  \\
    AnyEnhance     & 366.2M    & 1.300   \\
    \cdashline{1-3}\noalign{\vskip\belowrulesep}
    SenSE$_{tiny}$ & 248.5M    & 0.049  \\
    SenSE$_{base}$ & 1571.3M   & 1.122  \\
    \bottomrule
  \end{tabular}
  }
\end{table}

\begin{table}[t]
  \caption{Results of the proposed method and comparison models on multiple test sets. The boldface indicates the best result and the underline denotes the second best. `D' and `G' denote discriminative and generative methods, respectively. Reference speech is not used in the comparisons.}
  \label{tab:results-dns}
  \centering
  \resizebox{1.0\linewidth}{!}{%
  \setlength{\tabcolsep}{4pt}
  \begin{tabular}{lccccccc}
    \toprule
        \textbf{Model} 
        & \textbf{Type}
        & \textbf{DNSMOS}↑
        & \textbf{NISQA}↑ 
        & \makecell{\textbf{Speech-}\\\textbf{BERTScore}}↑ 
        & \textbf{dWER(\%)}↓ 
        & \textbf{SIM-o}↑ & 
    \\
    \midrule
    \multicolumn{7}{c}{\textbf{DNS Challenge \textit{no-reverb}}} \\
    \midrule
    Voicefixer            
                     & D   & 3.248  & 4.385  & 0.861  & 9.28  & 0.714   \\
    TF-GridNet            
                     & D   & 3.312  & 4.332  & 0.930  & \underline{4.22}  & \textbf{0.956}   \\
    PGUSE
                     & D+G & 3.333  & 4.661  & \underline{0.932}  & 4.40  & \underline{0.942}   \\
    \cdashline{1-8}\noalign{\vskip\belowrulesep}
    GenSE
                     & G   & \textbf{3.425}  & 4.672  & 0.838  & 20.08  & 0.285   \\
    FlowSE
                     & G   & 3.265  & 4.733  & 0.898  & 9.92  & 0.846   \\
    LLaSE-G1
                     & G   & \underline{3.415}  & 4.504  & 0.889  & 8.69  & 0.748 \\
    AnyEnhance       
                     & G   & 3.406  & \underline{4.784}  & 0.925  & 4.95  & 0.885   \\
    \cdashline{1-8}\noalign{\vskip\belowrulesep}
    SenSE$_{tiny}$     & G   & 3.375  & 4.757  & 0.929   & 4.85   & 0.890 \\
    SenSE$_{base}$     & G   & 3.376  & \textbf{4.788}  & \textbf{0.942}  & \textbf{3.89}  & 0.921   \\
    \midrule
    \multicolumn{7}{c}{\textbf{DNS Challenge HardSet}} \\
    \midrule
    Voicefixer        & D   & 3.119  & 3.958  & 0.779  & 27.01  & 0.553   \\
    TF-GridNet        & D   & 3.146  & 3.974  & 0.844  & 12.00  & 0.813   \\
    PGUSE             & D+G & 3.251  & 4.112  & 0.871  & 14.50  & \textbf{0.843}  \\
    \cdashline{1-8}\noalign{\vskip\belowrulesep}
    FlowSE            & G   & 2.940  & 4.057  & 0.799  & 29.38  & 0.684   \\
    LLaSE-G1          & G   & 3.370  & 4.366  & 0.830  & 28.47  & 0.604   \\
    AnyEnhance        & G   & 3.384  & \underline{4.817}  & 0.876  & 15.48  & 0.778   \\
    \cdashline{1-8}\noalign{\vskip\belowrulesep}
    SenSE$_{tiny}$    & G   & \underline{3.408}  & 4.624  & \underline{0.879}  & \underline{11.74}  & 0.792   \\
    SenSE$_{base}$    & G   & \textbf{3.408} & \textbf{4.873} & \textbf{0.915} & \textbf{8.93} & \underline{0.838} \\

    \midrule
    \multicolumn{7}{c}{\textbf{DNS Challenge GSR}} \\
    \midrule
    Voicefixer        & D   & 3.285 & 4.003 & 0.817 & 18.13 & 0.553 \\
    TF-GridNet        & D   & 3.233 & 4.041 & 0.872 & \underline{7.36}  & 0.767 \\
    PGUSE             & D+G & 3.291 & 4.166 & 0.896 & 8.52  & \textbf{0.805} \\
    \cdashline{1-7}\noalign{\vskip\belowrulesep}
    AnyEnhance        & G   & \textbf{3.396} & \textbf{4.847} & 0.896 & 12.25 & 0.779 \\
    \cdashline{1-8}\noalign{\vskip\belowrulesep}
    SenSE$_{tiny}$   & G   & 3.380 & 4.675  & \underline{0.908} & 8.10  & 0.783 \\
    SenSE$_{base}$    & G   & \underline{3.388} & \underline{4.806} & \textbf{0.922} & \textbf{6.70}  & \underline{0.797} \\

    \midrule
    \multicolumn{7}{c}{\textbf{VCTK GSR}} \\
    \midrule
    Voicefixer        & D   & 2.976  & 4.009  & 0.887  & 8.30  & 0.681   \\
    TF-GridNet        & D   & 3.054  & 4.425  & 0.924  & 3.77  & \underline{0.855}   \\
    PGUSE             & D+G & 3.057  & 4.419  & 0.928  & 3.10  & \textbf{0.879}   \\
    \cdashline{1-7}\noalign{\vskip\belowrulesep}
    AnyEnhance        & G   & \textbf{3.128}  & \textbf{4.756}  & 0.924  & 4.62  & 0.795   \\
    \cdashline{1-7}\noalign{\vskip\belowrulesep}
    SenSE$_{tiny}$    & G   & 3.106  & 4.584  & \underline{0.933}  & \underline{3.08}   & 0.797 \\
    SenSE$_{base}$    & G   & \underline{3.109}  & \underline{4.726}  & \textbf{0.943}  & \textbf{1.79}  & 0.824   \\

    \bottomrule
  \end{tabular}
  }
  
\end{table}

\subsection{Experimental Setup}

\textbf{Datasets}\quad We use clean speech data from the open-source Emilia ~\cite{he2024emilia} dataset to train our base and tiny model. We selected approximately 22,000 hours of data with DNSMOS scores greater than 3.40. Additionally, small models were trained for ablation studies using 1,400 hours of clean speech from the Interspeech 2020 DNS Challenge \cite{reddy2020interspeech} dataset. The noise datasets included DEMAND, ESC-50, and DNS Challenge \cite{cutler2024icassp}, totaling roughly 700 hours, while room impulse responses (RIRs) were sourced from openSLR26 and openSLR28 \cite{ko2017study}. Paired training data are simulated using the parameter settings summarized in \Cref{tab:distortion}.

For speech denoising, we evaluate on the official DNS Challenge no-reverb test set. To assess performance under low signal-to-noise ratio (SNR) conditions, we additionally construct the DNS Challenge HardSet by re-mixing speech and noise from the no-reverb test set with SNRs randomly sampled from (-5, 0) dB. To evaluate robustness across diverse distortion types, we further construct two simulated General Speech Restoration (GSR) test sets, namely DNS Challenge GSR and VCTK GSR. In DNS Challenge GSR, clean speech is processed using our simulation pipeline and then mixed with noise, while in VCTK GSR, clean speech from the VCTK dataset is similarly processed and mixed with unseen noise and room impulse responses (RIRs), resulting in 167 test samples.

\textbf{Model Details}\quad In this paper, we employ three model variants with different scales. The base model SenSE$_{base}$ is used to explore the upper bound of performance of the proposed approach, the small model SenSE$_{small}$ is adopted for ablation studies, and the tiny model SenSE$_{tiny}$ is designed to evaluate performance under low computational budgets. Notably, the tiny model adopts an encoder-only language model, enabling single-pass inference.
The detailed model configurations are summarized in \Cref{tab:model_config}, where the entries in the SASLM and DiT Blocks columns indicate the hidden size, number of layers, and number of attention heads of the corresponding Transformer modules, respectively. For the Whisper encoder and the vocoder, we use the official open-source pre-trained models. During training, the masking ratios for clean and degraded speech are randomly sampled from the ranges of (70\%,100\%) and (50\%,100\%), respectively.

For the base and tiny model, the SASLM was trained on 8 NVIDIA RTX 5880 Ada Generation GPUs with a batch size of 76,800 audio frames for 1.15M steps. During the first 550K steps, the audio encoder was frozen while only the language model was optimized; in the subsequent steps, the language model was frozen, and the audio encoder was fine-tuned. The flow matching stage was trained on the same GPUs with a batch size of 153,600 audio frames for 350K steps. For the small model, both stages were trained on 8 NVIDIA RTX 4090 GPUs with a batch size of 147,200 audio frames for 300K steps. Across all training configurations, the AdamW optimizer was employed with a peak learning rate of 7.5e-5, using 20k linear warm-up steps followed by linear decay.

\begin{figure*}[t]
    \centering
    \begin{subfigure}{0.32\linewidth}
        \includegraphics[width=\linewidth]{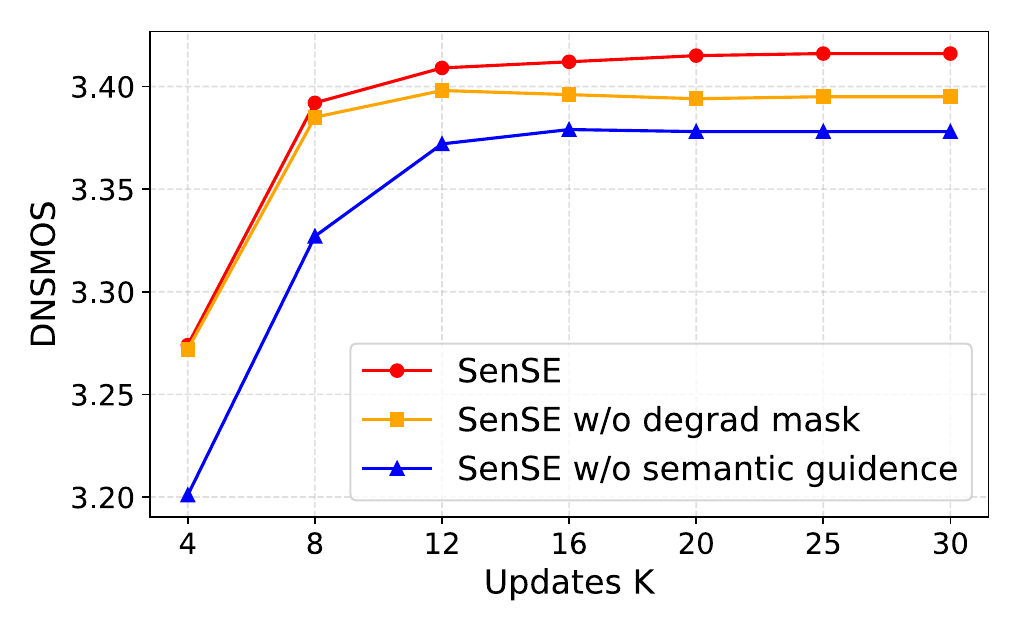}
    \end{subfigure}
    \hfill
    \begin{subfigure}{0.32\linewidth}
        \includegraphics[width=\linewidth]{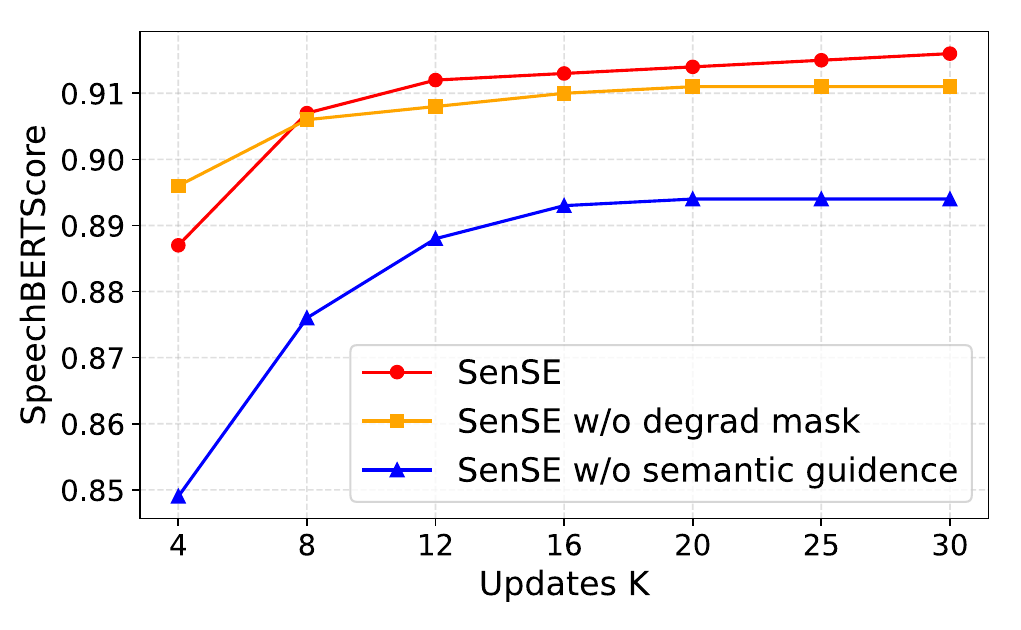}
    \end{subfigure}
    \hfill
    \begin{subfigure}{0.32\linewidth}
        \includegraphics[width=\linewidth]{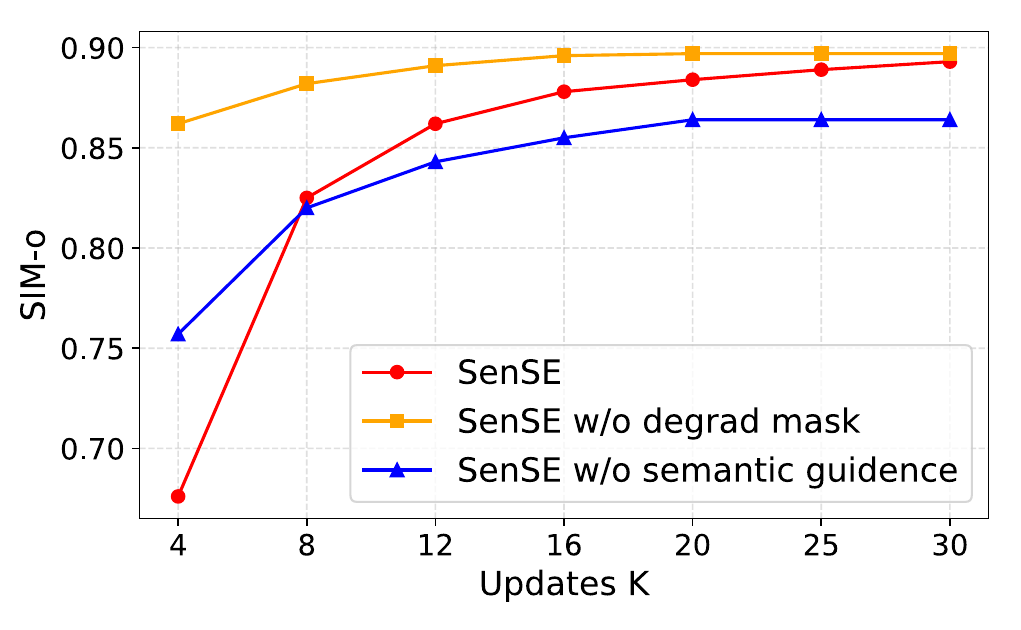}
    \end{subfigure}
    
    \caption{Results of the ablation study on the SenSE framework, where "w/o" indicates the removal of a specific method or component.}
    \label{fig:ablation}
\end{figure*}

\textbf{Comparison Models}\quad We compare our model against state-of-the-art speech enhancement systems, including both discriminative models: VoiceFixer ~\cite{liu2022voicefixer}, TF-GridNet ~\cite{wang2023tf}; as well as generative models: GenSE ~\cite{yao2025gense}, AnyEnhance ~\cite{zhang2025anyenhance}, LLaSE-G1 ~\cite{kang2025llase}, and FlowSE ~\cite{wang2025flowse}; as well as the hybrid predictive-generative model PGUSE ~\cite{zhang2025composite}. For AnyEnhance, we adopt the official implementation without prompt-guidance and self-critic modules, in order to isolate and evaluate the impact of the model architecture itself. The model is retrained using the same training data as SenSE$_{base}$ to ensure a fair comparison. For VoiceFixer, GenSE,  LLaSE-G1, and FlowSE, we directly use the officially released pretrained models. For TF-GridNet and PGUSE, we retrain the models using 1,400 hours of clean speech from the DNS Challenge, combined with the same noise and RIR dataset used for training SenSE$_{base}$.

\textbf{Metrics}\quad We evaluate our model using several commonly adopted metrics for speech enhancement. \textbf{DNSMOS:} a reference-free perceptual quality metric ~\cite{reddy2021dnsmos}. We use the OVAL score to evaluate speech quality. \textbf{NISQA:} a reference-free perceptual quality metric designed for 48 kHz speech ~\cite{mittag2021nisqa}. \textbf{SpeechBERTScore:} a metric for assessing the semantic similarity between enhanced and reference speech ~\cite{saeki2024speechbertscore}. In this work, we use HuBERT-base\footnote{https://huggingface.co/facebook/hubert-base-ls960.} model to extract the feature. \textbf{dWER:} a metric that evaluates the word error rate (WER) difference between enhanced and reference speech using ASR transcriptions. We adopt Whisper-large-v3 ~\cite{radford2023robust} as the ASR model for this evaluation. \textbf{SIM-o:} a measure of timbre similarity between enhanced and reference speech ~\cite{le2023voicebox}. In this work, we compute cosine similarity between speaker embeddings extracted using a WavLM-large-based speaker verification model \cite{chen2022large}.

\begin{table}[t]
  \caption{Effect of reference speech.}
  \label{tab:results-prompt}
  \centering
  \resizebox{0.95\linewidth}{!}{%
  \setlength{\tabcolsep}{4pt}
  \begin{tabular}{lcccccc}
    \toprule
        \textbf{Model} 
        & \textbf{DNSMOS}↑ 
        & \textbf{NISQA}↑  
        & \makecell{\textbf{Speech-}\\\textbf{BERTScore}}↑ 
        & \textbf{dWER(\%)}↓ 
        & \textbf{SIM-o}↑ &
    \\

    \midrule

    SenSE                & \textbf{3.109}  & 4.726  & 0.943  & \textbf{1.79}  & 0.824   \\
    SenSE(w/ reference)  & 3.108  & \textbf{4.728} & \textbf{0.947} & 1.84 & \textbf{0.873} \\
    
    \bottomrule
  \end{tabular}
  }
\end{table}

\subsection{Experimental Results}

\Cref{tab:results-dns} present the main results on speech denoising and the general speech restoration tasks. During inference of SenSE, sampling in SASLM is disabled to ensure relatively stable outputs. We report the average score of SenSE and the sampling-based baselines using five different random seeds. By default, we set the CFG strength ~\cite{ho2022classifier} to 0.5, the sway sampling coefficient ~\cite{chen2024f5} to -1, and the number of function evaluations (NFE) to 8 for our flow matching model. Note that all enhanced outputs from the models are resampled to 16 kHz.

Table 3 reports the parameter counts and real-time factors (RTF) of SenSE$_{tiny}$, SenSE$_{base}$, and other baselines, where RTF is computed based on the inference time for 10-second audio samples. SenSE$_{tiny}$ achieves lower parameter count and RTF than all other generative baselines, enabling evaluation of performance under low computational complexity. By scaling up the model size, SenSE$_{base}$ is used to assess the upper performance bound of the proposed framework.

We draw the following conclusions from the experimental results: 1) On speech quality metrics (DNSMOS and NISQA), SenSE$_{base}$ achieves performance comparable to other generative approaches, being only slightly inferior to AnyEnhance in a few cases, while exhibiting clear advantages over discriminative methods; 2) On semantic similarity metrics (SpeechBERTScore and dWER), SenSE$_{base}$ consistently outperforms all baseline models, demonstrating its effectiveness in alleviating semantic inconsistency in generative speech enhancement and highlighting its potential to surpass discriminative approaches; 3) In terms of speaker similarity (SIM-o), SenSE$_{base}$ surpasses all generative baselines and achieves performance that is highly competitive with discriminative methods; 4) Even with a substantially reduced model size, SenSE$_{tiny}$ still significantly outperforms other generative approaches on most metrics, falling slightly behind AnyEnhance only in speech quality scores. Given SenSE’s superior ability to preserve speech content, this performance gap is considered acceptable; 5) Results on the DNS Challenge HardSet further indicate that scaling up SenSE yields substantial performance gains under severe distortion conditions, underscoring its strong performance ceiling and scalability.

We evaluate the effect of reference speech on the VCTK GSR test set. For each test sample, a different utterance from the same speaker is selected as the reference. As shown in \Cref{tab:results-prompt}, incorporating reference speech leads to a substantial improvement in speaker similarity, while achieving comparable or slightly improved performance on the other evaluation metrics compared to the reference-free setting.

\subsection{Ablation Studies and Analysis}

To validate the effectiveness of the proposed degrad mask and semantic guidance mechanism in SenSE, we conduct ablation experiments. Specifically, 1) we removed the mask applied to degraded speech in the flow matching stage; 2) we removed the first-stage SASLM and eliminated the semantic token from the conditions in the second-stage flow matching module.

The ablation study results are presented in \Cref{fig:ablation}. We summarize our findings as follows: 1) The semantic guidance mechanism consistently improves performance across DNSMOS, SpeechBERTScore, and SIM-o, indicating that incorporating additional semantic information effectively enhances the performance of flow matching-based speech enhancement models, particularly in terms of semantic preservation; 2) Introducing the degrad mask leads to inferior performance during the early stages of training. However, as training progresses, the model equipped with degrad mask surpasses its counterpart without degrad mask on DNSMOS and SpeechBERTScore, while the performance gap on SIM-o gradually narrows. Overall, the degrad mask yields a positive and stable impact on model performance.

\section{Conclusion}

In this study, we propose SenSE, a novel two-stage speech enhancement framework that incorporates language-model-derived semantic tokens to guide flow matching-based generation, effectively mitigating semantic inconsistency. A dual-path masked conditioning training strategy is further introduced to enhance semantic guidance and enable the use of prompt speech for recovering speaker information. We will release our code and models to facilitate reproducibility and further research in this field.

\bibliographystyle{IEEEbib}
\bibliography{icme2026references}

\end{document}